\documentclass[aps,prb,twocolumn,superscriptaddress,showpacs,floatfix]{revtex4-2}
\usepackage{comment}
\usepackage{physics}
\usepackage{amsmath,amsthm,amssymb,amsfonts, color, comment, graphicx}
\usepackage{xcolor}
\usepackage{float}
\usepackage{graphicx}
\usepackage{dcolumn}
\usepackage{bm}
\UseRawInputEncoding 
\usepackage{dsfont}
\usepackage{hyperref}

\newcommand{\mfp}{\mathfrak{p}}
\newcommand{\mfq}{\mathfrak{q}}
\newcommand{\mfh}{\mathfrak{h}}

\newcommand{\be}{\begin{equation}}
\newcommand{\ee}{\end{equation}}

\newcommand{\br}{{\bf{r}}}

\newcommand{\beal}{\begin{align}}
\newcommand{\eeal}{\end{align}}
\newcommand{\ra}{\rangle}
\newcommand{\la}{\langle}

\def\i{i}

\usepackage{graphicx}
\usepackage[caption=false]{subfig}

\begin{document}

\preprint{APS/123-QED}
\title{Phonon-driven multipolar dynamics in a spin-orbit coupled Mott insulator}
\author{Kathleen Hart}
\thanks{These authors contributed equally to this work.}
\affiliation{Department of Physics, University of Toronto, 60 St. George Street, Toronto, ON, M5S 1A7 Canada}
\author{Ruairidh Sutcliffe}
\thanks{These authors contributed equally to this work.}
\affiliation{Department of Physics, University of Toronto, 60 St. George Street, Toronto, ON, M5S 1A7 Canada}
\author{Gil Refael}
\affiliation{Department of Physics, California Institute of Technology, Pasadena CA 91125, USA}
\affiliation{Institute for Quantum Information and Matter, California Institute of Technology, Pasadena CA 91125, USA}
\author{Arun Paramekanti}
\email{arun.paramekanti@utoronto.ca}
\affiliation{Department of Physics, University of Toronto, 60 St. George Street, Toronto, ON, M5S 1A7 Canada}

\date{\today}

\begin{abstract}
Motivated by advances in pump-probe experiments and light-driven phenomena, we theoretically study the 
impact of pumped and driven 
phonons in Mott insulators which host multipole moments, thus going beyond conventional dipolar magnetism. 
As a case study, we examine pseudospin-1/2 Mott insulators
hosting quadrupolar and octupolar moments, and investigate the effect of resonantly exciting 
${\cal E}_g$ phonon modes which couple linearly 
to the quadrupoles. We show that this leads to multipolar precession,
with the backaction resulting in pseudochiral phonon dynamics in the octupolar ordered phase. We 
further study the impact of a short-time
two-phonon drive, showing that it can induce the build-up of octupolar order or even switch its sign on 
picosecond timescales. Our results are obtained using
a Monte Carlo code incorporating phonons, molecular dynamics simulations to numerically integrate the 
coupled spin-phonon equations of motion, and analytical Floquet theory. Our work shows how 
driven phonons can probe and control hidden orders in solids.
\end{abstract}

\maketitle

\noindent {\it Introduction.---}
Shaking electrons or atoms with light can induce remarkable
non-equilibrium many body states in solid state crystals 
as well as ultracold atoms \cite{Intro_Floquet_TI_Refael_NatPhys2011,Intro_floquet_Lindner_PRB2013,bukov2015universal,Intro_Floquet_Eckardt_RMP2017,Intro_Floquet_Oka_ARCMP2019,Intro_Floquet_Rudner_NatRevPhys2020,Intro_Floquet_Fiete_AnnPhys2021,junk2020floquet}.
In recent years, rapid advances in 
ultrafast and THz optics have led to the ability to resonantly excite specific orbital transitions and phonon modes in solids
\cite{Intro_NonlinearPhononics_Cavalleri_NatPhys2011,Afanasiev_Science2021,Intro_Cavalleri_nonlinearTHZ,liu2018floquet}, providing a
distinct pathway to control
broken symmetry states including ferroelectricity
\cite{Intro_Cavalleri_ferroelec_PRL2017}, superconductivity \cite{Intro_Cavalleri_HiTcPhonon_PRX2020}, and 
magnetism \cite{Intro_OpticalCEF_Cavalleri_NatPhy2020,Afanasiev_Science2021,Intro_YTiO3FM_Cavalleri_Nature2023,Mathiessen_PRL2023,afanasiev2021ultrafast,stupakiewicz2021ultrafast}.
In dipolar magnets, Einstein phonons can strongly 
impact the spin-spin exchange interaction, stabilizing unconventional 
equilibrium magnetic orders in geometrically
frustrated lattices such as kagome, triangular, or pyrochlore systems \cite{bergman_prb2006,wang2008spin,szasz2022phase,watanabe2023magnetic}.
In addition to equilibrium orders, recent experimental and theoretical
work has shown that coherent phonon driving can produce fascinating non-equilibrium Floquet dynamics in spin-phonon coupled
dipolar magnets. These include anomalously large effective magnetic fields from driven chiral phonons 
\cite{nova2017effective,juraschek2017dynamical,geilhufe2021dynamically,juraschek2022giant,luo2023large,davies2024phononic,basini2024terahertz}, tuning of magnetic orders and magnon band topology \cite{Intro_Fiete_CrI3FMAFM_PRB2020, Afanasiev_Science2021,Intro_YTiO3FM_Cavalleri_Nature2023,phonon_ferromagnetism_narang_prb2023},
frustration breaking in dimerized magnets \cite{Intro_ShastrySutherland_Ruegg_PRB2023}, 
nonequilibrium phase transitions 
\cite{spindyn_yarmohammadi2023laserenhanced}, and high harmonic generation \cite{spindyn_hhg_allafi2024spin}.
In addition, this has motivated the study of
many-body spin-phonon dynamics and
Floquet-phonon physics in layered van der Waals materials \cite{Shin2018,Hubener2018,Padmanabhan_NatComms2022,nuske2020floquet}.

Going beyond dipolar spin systems, there is a rich set of issues one encounters for
\emph{multipolar} degrees of freedom which are found in solids with strong spin-orbit coupling. Unlike dipole moments, 
higher order multipoles such as quadrupoles and
octupoles are associated with spin-orbital entangled charge or current distributions which do not simply couple to conventional 
probes; for this reason, multipolar broken symmetries, found in diverse materials, 
are hard to detect and generically termed `hidden orders' \cite{santini_RMP2009,santini2000magnetic,tokunaga2006nmr,chandra2002hidden,tripathi2007sleuthing,rau2012hidden,Intro_Review_HiddenOrder_Mydosh2020,sakai2011kondo,sato2012ferroquadrupolar,lee2018landau,patri2019unveiling,fu2015parity,harter2017parity,voleti2023probing}.
Quadrupolar moments, however, being even under time-reversal symmetry, can {\it linearly}
couple to phonons, leading to significant differences in the equilibrium ordering and excitations of multipoles
when compared with magnetic dipoles. This is exemplified by recent discoveries of unusual multipolar symmetry breaking 
and lattice displacements in Ba$_2$MgReO$_6$ \cite{hirai2019successive,streltsov2020jahn,hirai2020detection,mansouri2021untangling,mosca2024interplay} 
and the detection of vibronic modes via resonant inelastic X-ray scattering in 
K$_2$IrCl$_6$ \cite{warzanowski2024spin}. 

Recent experiments have begun to extend optical and THz techniques 
to study the non-equilibrium dynamics of multipolar orders. For instance, resonantly excited phonon modes in pump-probe and THz 
experiments have shown evidence of multipolar modes coupled to lattice vibrations \cite{Intro_ShastrySutherland_Ruegg_PRB2023}. 
In $\rm{Ca_2RuO_4}$, THz experiments have revealed a `hidden' quadrupolar-ordered state through the anomalous broadening of coherently driven 
phonons \cite{ning2023coherent}. 
These developments call out for a theory of many-body dynamics in
pumped and driven multipolar quantum materials as a route to probe and control 
hidden orders.


In this Letter, we fill this important gap by developing a simulation tool to study 
multipolar orders and their semiclassical dynamics including the crucial effect of 
coupling to Einstein phonons with
dissipation. This method combines large scale equilibrium spin-phonon Monte Carlo (MC) simulations with spin-phonon 
molecular dynamics (MD) obtained by integrating their coupled equations of motion. 
The key idea underlying our work is that a phonon mode coupled to a quadrupole acts as a transverse field to all other orthogonal multipoles, just as for a dipolar spin, a magnetic field along $S_z$ acts as a transverse field to $(S_x,S_y)$. For multipolar materials,
kicking the phonon thus engenders nontrivial precessional dynamics in multipolar space which we leverage as a route to unearth hidden orders.

While our approach is broadly applicable to
many multipolar materials, we illustrate the power of this tool
by focusing on a recently discovered paradigmatic example in Mott 
insulators, the osmate double perovskites \cite{maharaj2020octupolar,voleti2020multipolar,voleti2021octupolar,voleti2023probing,khaliullin2021exchange,churchill2022competing}
which display competing octupolar and quadrupolar ordering tendencies \cite{voleti2020multipolar,voleti2021octupolar,churchill2022competing}.
Our work shows how the presence of phonons, 
resonantly excited by either pump pulses or Floquet driving, can influence 
the dynamics of the multipolar moments and enable the detection, build-up, and ultrafast switching of exotic octupolar order. 

\noindent {\it Model.---}
The simplest example of multipolar moments arise in materials with non-Kramers doublets, local pseudospin-1/2 degrees of freedom
whose doublet degeneracy is protected by crystalline point group symmetries rather than just time-reversal symmetry as in the
case of Kramers doublets. 
Remarkably, non-Kramers doublets are found in a
diverse array of quantum materials including: quantum spin-ice compounds such as {Pr$_2$Zr$_2$O$_7$}, {Pr$_2$Sn$_2$O$_7$} \cite{matsuhira2009spin,lee2012generic,kimura2013quantum,patri2020theory},
Kondo lattice intermetallics {PrTi$_2$Al$_{20}$, {PrV$_2$Al$_{20}$}} \cite{sakai2012superconductivity,taniguchi2016nmr,lee2018landau,freyer2018two,patri2019unveiling},
Mott insulating double perovskites $\rm{Ba_2CaOsO_6}$, $\rm{Ba_2MgOsO_6}$, $\rm{Ba_2ZnOsO_6}$ \cite{maharaj2020octupolar,voleti2020multipolar,voleti2023probing},
and vacancy ordered double perovskites {Cs$_2$WCl$_6$} and {Rb$_2$WCl$_6$} \cite{pradhan2024multipolar,morgan2023hybrid}.

We focus here on non-Kramers doublets with time-reversal even (${\cal T}$-even)
quadrupolar moments represented by pseudospin-1/2 Pauli operators $(\tau_x,\tau_z)$
and a ${\cal T}$-odd octupole represented by $\tau_y$ as realized in the double perovskites and the Kondo intermetallics. 
In the double perovskites, these
pseudospins live on a face-centered
cubic (fcc) lattice (see End Matter), and descend from splitting a $J=2$
angular momentum multiplet, with
$\tau_x \propto (J_x^2-J_y^2)$, $\tau_z \propto (3 J_z^2-J(J+1))$, and
$\tau_y \propto \overline{J_x J_y J_z}$, with bar denoting operator symmetrization.

Symmetries dictate the following fcc lattice Hamiltonian describing nearest-neighbor pseudospin interactions
\begin{eqnarray}
\label{eq:nk1}
\!\! \mfh_{\rm sp} \!&=&\!\!\! \sum_{\langle \br,\br'\rangle}\!\!\! \left[- J_0 \tau_{\br y} \tau_{\br' y}
\!+\! J_1 \left(\cos^2\!\phi_{\br\br'} \!+\!  \gamma  \sin^2\!\phi_{\br\br'} \right) \tau_{\br z} \tau_{\br' z} \right. \nonumber \\ 
 &+& J_1 \left( 1-\gamma \right) \sin\phi_{\br\br'} \cos\phi_{\br\br'} \left( \tau_{\br x}\tau_{\br' z} + \tau_{\br z}\tau_{\br' x} \right) \nonumber \\
 &+&  \left. J_1 \left(\sin^2\phi_{\br\br'} + \gamma \cos^2\phi_{\br\br'}  \right)\tau_{\br x} \tau_{\br' x} \right],
\end{eqnarray}
where $\phi_{ij} = \{ 0 , 2\pi/3 , 4\pi/3 \}$ correspond to nearest neighbors $\la\br\br'\ra$ in the $\{ XY,YZ,ZX \}$ planes. 
The phase diagram of this model
depends on the dimensionless ratios of exchange couplings via
$J_1/J_0$ and $\gamma$ \cite{paramekanti2020octupolar,voleti2021octupolar}.

Based on density functional theory (DFT) and dynamical mean field theory for Ba$_2$CoOsO$_6$ 
\cite{pourovskii2021ferro}, we set $\gamma\! \approx\! -0.4$ and $J_1/J_0 \! \approx\! 0.5$
for the pseudospin Hamiltonian; 
this places us in the ferro-octupolar phase for $T/J_0 \lesssim 3.2$ as determined from
classical Monte Carlo simulations \cite{voleti2021octupolar}. A microscopic calculation of
the octupolar exchange yields $J_0 \approx 1$\,meV, which leads to $T_{\rm oct} \approx 40$\,K, in good agreement
with experiments on Ba$_2$CoOsO$_6$ \cite{maharaj2020octupolar,voleti2021octupolar,voleti2023probing}. To
construct a general phase diagram, we fix $\gamma$ and vary
$J_1/J_0$ and $T/J_0$ to access both ferro-octupolar and quadrupolar ordered phases.

Next, we consider cubic symmetry ${\cal E}_g$ Einstein phonon modes at each lattice site
with $x \equiv x^2\!-\!y^2$, $z \equiv 3z^2\!-\! r^2$, with the phonon Hamiltonian
\begin{equation}
\label{eq:nk2}
    \mfh_{\rm{ph}}= \sum_{\br,\alpha=x,z} \left[ \frac{1}{2 m} \mfp_{\alpha}^2(\br)  + \frac{1}{2} m \omega^2 \mfq_{\alpha}^2(\br) \right]
\end{equation}
where $(\mfq_\alpha,\mfp_\alpha)$ are the normal mode coordinates and momenta, and $\omega,m$ represent the frequency and mass, respectively. 
On symmetry grounds, the ${\cal T}$-odd 
octupole will not linearly couple to phonons, however the $(\tau_x, \tau_z)$ quadrupoles can couple linearly
to the phonon modes. The symmetry allowed spin-phonon coupling is given by
\begin{equation}
\label{eq:nk3}
    \mfh_{\rm sp-ph} = - g  \sum_\br \left[ \mfq_x(\br) \tau_{\br x} + \mfq_z(\br) \tau_{\br z} \right]
\end{equation}
where $g$ is the strength of the spin-phonon coupling.

\begin{figure*}[t]
\includegraphics[width=1.0\textwidth]
{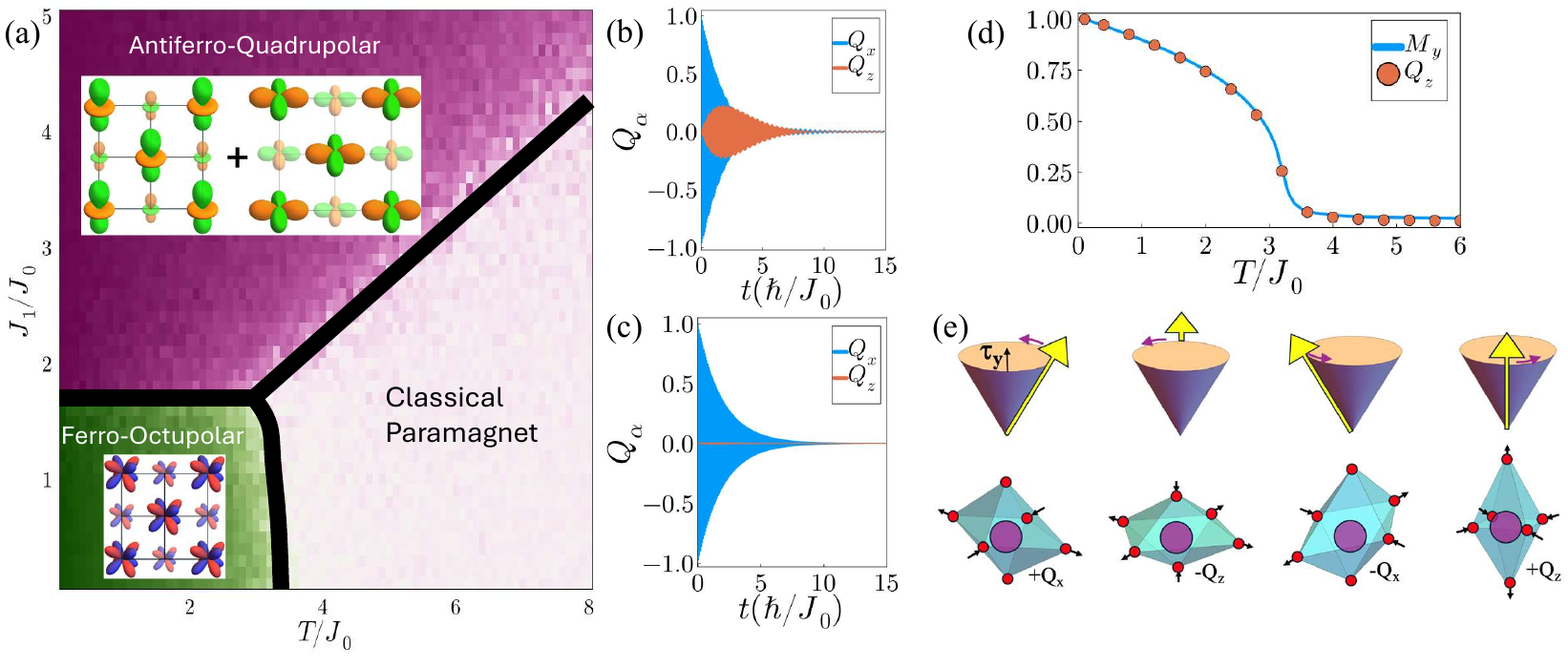}
\caption{ 
(a) Monte Carlo phase diagram of the Hamiltonian in Eqs. \ref{eq:nk1}, \ref{eq:nk2} and \ref{eq:nk3} as a function of $T/J_0$ and $J_1/J_0$
(for fixed $\gamma=-0.4$). Colors in the ordered phases depict strength of the ferro-octupolar order $\langle\tau_y\rangle$ (green)
and antiferro-quadrupolar (AFQ) order (purple); insets depict the AFQ and ferro-octupolar orders on two adjacent layers of the fcc lattice 
(top layer: bold/large, bottom layer: light/small) with orbital shapes and colors respectively showing the charge density and the sign,
and the `+' for AFQ indicating a superposition of the two orbital orders.
(b) and (c) Phonon dynamics in pump-probe simulations induced by initial momentum kick for $Q_x$ phonon 
for $J_1/J_0=1.3$ and phonon damping constant $2\pi\eta/\omega=0.07$ \cite{maehrlein2017terahertz,juraschek2018sum}, averaged over 160 different initial conditions. 
(b) $T=0.1 J_0 < T_{\rm oct}$, single domain ferro-octupolar; (c) $T = 5.2 J_0 > T_{\rm{oct}}$, paramagnet.  
For $T<T_{\rm{oct}}$, we observe coherent energy transfer between $Q_x$ and $Q_z$ phonons, which is absent 
for $T>T_{\rm{oct}}$. (d) Comparison showing qualitative agreement between 
octupolar order parameter $M_y = \langle \tau_y \rangle$ (blue dots) and $Q_z$-$Q_x$ cross-amplitude, i.e. maximum $Q_z$ response 
to $Q_x$ pump (orange dots) as shown in panels (b) and (c); both curves are scaled by their lowest $T$ data point.  (e) Schematic of 
energy exchange between the $(Q_x,Q_z)$ phonons in the ferro-octupolar phase through pseudospin precession and quadrupole-phonon 
coupling. }

\label{fig:PD_octupolar}
\end{figure*}

\noindent {\it Equilibrium phase diagram.---}
We begin by constructing 
the equilibrium phase diagram of the full model $H=H_{\rm sp}+H_{\rm ph}+H_{\rm sp-ph}$
in the presence of coupling to the cubic ${\cal E}_g$ phonon modes.
For this purpose, it is useful to define a length scale $a\!=\!0.01 a_0$, where 
$a_0 \!\approx\! 4$\,{\AA} is the cubic lattice constant \cite{wakeshima2013physical}, and
a normalized phonon amplitude $Q_\alpha = \mfq_\alpha/a$;
in these units, $Q_\alpha =1$
implies an experimentally accessible $1\%$ atomic displacement, relative to the lattice constant \cite{Intro_OpticalCEF_Cavalleri_NatPhy2020}.
The normalized conjugate momenta are $P_\alpha = a \mfp_\alpha$.
Measuring energy in units of $J_0$ yields dimensionless Hamiltonians,
\begin{eqnarray}
H_{\rm sp} &=& \mfh_{\rm sp}/J_0 \\
H_{\rm ph} &=& \!\!\! \sum_{\br,\alpha=x,z} \left[\frac{P_{\alpha}^2(\br)}{2 M} + \frac{1}{2} M \Omega^2 
Q_{\alpha}^2(\br) \right] \label{eq:hph},\\
H_{\rm sp-ph} &=& - \lambda \sum_\br \left[ Q_x(\br) \tau_{\br x} + Q_z(\br) \tau_{\br z} \right] \label{eq:hspph}
\end{eqnarray}
where the dimensionless coefficients in Eqn.~\ref{eq:hph}, \ref{eq:hspph} are
$1/2M=\hbar^2/2 m J_0 a^2$,
$\Omega= \hbar\omega/J_0$, and
$\lambda= g a/J_0$.
This choice of energy units
corresponds to measuring time in units of $\hbar/J_0$ ($\approx\! 0.6$ picosecond).
We set $m = m_{\rm oxygen} \approx 2.6 \times 10^{-26} \rm kg$, since we are simulating ${\cal E}_g$ phonon
modes where only the oxygen atoms have finite displacements,
and we set
$\hbar\omega \approx 80$\,meV as a typical ${\cal E}_g$ phonon mode frequency; using this,
$1/2M \approx 80$, and $\Omega \approx 80$.
From DFT, a $\sim\!1\%$ atomic displacement leads to a $\sim\!5$\,meV splitting of the 
non-Kramers doublet \cite{voleti2023probing}; this implies  $g=5 {\rm meV}/a$, 
so the dimensionless spin-phonon coupling
$\lambda \approx 5$. 

Fig.~\ref{fig:PD_octupolar}(a) displays the phase diagram obtained from our MC simulations (on system size
with $10^3$ sites) as we
vary the temperature $T/J_0$ and quadrupolar exchange $J_1/J_0$, effectively sampling spin and phonon configurations from thermal populations. We find
a high temperature paramagnetic phase, a low temperature ferro-octupolar phase for $J_1/J_0 \! \lesssim \! 1.8$
and an antiferro-quadrupolar (AFQ) phase with wavevector $(\pi,\pi,0)$ (and equivalent) for $J_1/J_0 \! \gtrsim \! 1.8$. At low $T$,
phonon displacements are absent in the ferro-octupolar phase, while the AFQ phase shows small
displacements $Q_\alpha \!\sim\! 0.1$ which track the local quadrupole order. While $ Q_\alpha(\br) \neq 0$ 
at each site, the sign of $Q_\alpha(\br)$ alternates from layer to layer (see Fig.~\ref{fig:PD_octupolar}). 
Comparable staggered lattice displacements,
$\sim \! 0.1\%$,
have been found in the AFQ phase of the $j=3/2$ Mott insulator Ba$_2$MgReO$_6$ \cite{hirai2020detection,mosca2024interplay,hirai2019successive}.
We next explore the out-of-equilibrium pseudospin-phonon dynamics.

\noindent {\it Detection of octupolar order. ---}
In pump-probe experiments \cite{ning2023coherent}, an incoming pump beam can excite the even-parity cubic
 ${\cal E}_g$ phonon modes via a two-photon Raman
transition or via non-linear phononics, with the subsequent dynamics providing useful information about broken symmetries. 
To simulate
pump-probe dynamics, we use equilibriated MC spin-phonon
configurations as
initial conditions, impulsively excite the $Q_x$ phonon, 
and time-evolve the coupled spin and phonon degrees of freedom;
see End Matter for simulation details.

Fig.~\ref{fig:PD_octupolar}(b)
shows the time evolution of the phonon displacements $Q_x(t)$ and $Q_z(t)$ when we give an impulsive kick to the $Q_x$ phonon at time $t=0$
in the ferro-octupolar and paramagnetic phases. The rapid oscillations of the $Q_x, Q_z$ modes at the phonon frequency are not visible
in the plot, but are shown in the Supplementary Material (SM) \cite{suppmat}.
Remarkably, we see that octupolar broken symmetry  for $T < T_{\rm oct}$
mediates coherent energy transfer from the $Q_x$ to $Q_z$ mode on a time scale $t\!\approx\! 5 \hbar/J_0$ ($\approx 3$\,ps).
Such energy transfer is absent due to rapid dephasing in the paramagnet ($T \!>\! T_{\rm oct}$) 
or in the quadrupolar phase; this is shown in Fig.~\ref{fig:PD_octupolar}(c) for the paramagnet.


As shown in Fig.~\ref{fig:PD_octupolar}(d), the maximum cross-amplitude, i.e. the highest
amplitude of $Q_z$ oscillations for an impulsive kick to the $Q_x$ phonon, serves as a direct probe of the octupolar order
parameter. We understand this as follows (see also SM \cite{suppmat}).
The octupolar order leads to an average internal field 
along $\tau_y$, while the pumped phonon tips the initial pseudospin state away from $\tau_y$. This leads 
to coherent
pseudospin precession around the $\tau_y$-axis, serving as a conduit to transfer energy from $Q_x$ to $Q_z$ and back; see Fig.~\ref{fig:PD_octupolar}(e).
This protocol could be realized using a stimulated Raman scattering mechanism
\cite{de1985femtosecond} to pump one phonon mode
followed by time-resolved X-ray diffraction (trXRD)
to probe time-dependent atomic positions for extracting the amplitudes of both phonon modes as a function of time delay. 
Ref. \cite{gerber2017femtosecond} identifies deformations of order $\sim 3\times 10^{-2} ${\AA} for a bond of length $\sim 2.67${\AA}, 
making feasible the detection of the $\sim1\%$ distortions discussed in our work.
The maximum amplitude of the $Q_z$ phonon mode, a signature of octupolar order,
is revealed on timescales $t \sim 2 \hbar/J_0$ (about $1.3 \rm ps$), 
which is reasonable for trXRD experiments.



\begin{figure}[t]
\centering
\includegraphics[width=0.48\textwidth]{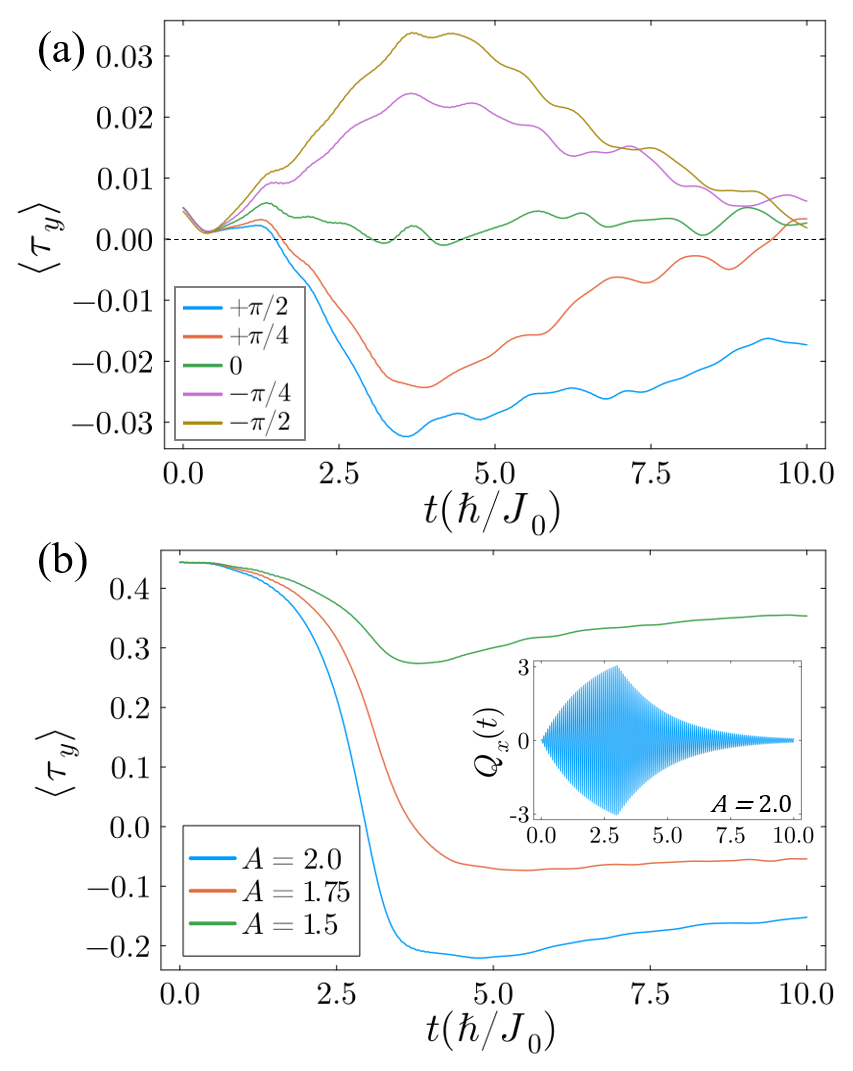}
\caption{Dynamics of $\langle\tau_y\rangle$ in the presence of short-time phase coherent two-mode phonon drive with weak dissipation.
The drive $H_{\rm drive}^{\rm ph}$ is turned on from $t=0\hbar/J_0$ to $t=3\hbar/J_0$ and then turned off.
(a) In the paramagnet $T \!=\! 1.1 T_{\rm oct}$, with drive strength $A\!=\!0.5$ (peak phonon
amplitude $Q_0 \!\approx\! 0.75$) and various
phase differences $\phi \in [-\pi/2,\pi/2]$, showing that driven phonons can select the sign of generated octupolar order.
(b) In the ferro-octupolar ordered phase $T \!=\!0.8 T_{\rm oct}$, with varying drive strength $A$, which induces a peak
phonon amplitude $Q_0$ between $2-3$. 
With a suitable phase choice $\phi\!=\!\pi/2$, it can be used to decrease the  octupolar order parameter $\langle\tau_y\rangle$, and even switch from positive to negative. Inset: Phonon coordinate $Q_x(t)$ for $A\!=\!2$, showing its amplitude build-up
and eventual dissipation after the drive is switched off.}
\label{fig:nk2phonon}
\end{figure}

\noindent {\it Dynamical build-up of octupolar order. ---}
We next explore how a simultaneous drive of both ${\cal E}_g$ phonon modes
can be used to control octupolar ordering.
We consider $H_{\rm drive}^{\rm ph} \!=\! - A(t) \sum_\br [Q_x(\br) \cos (\Omega t) \!+\! Q_z(\br) \cos (\Omega t - \phi)]$
which describes a coherent drive of Raman-active phonons as a result of terahertz sum-frequency excitation; the drive strength $A(t)$ is thus proportional to the square of the electric field \cite{maehrlein2017terahertz,juraschek2018sum}. We implement a short-pulse drive amplitude $A(t)\! = \! A$ 
for $t \!<\! 3\hbar/J_0$ and $A(t)\!=\! 0$ otherwise.
Fig.~\ref{fig:nk2phonon}(a)
shows the effect of this drive in the paramagnetic phase,
averaged over 320 initial MC configurations, for phases  
$\phi=(0,\pm\pi/4,\pm\pi/2)$. For a drive strength  which leads to a peak phonon amplitude $Q_0 \sim 0.75$, 
we find that, depending on the relative phase, the drive pulse can induce transient octupolar
order $\langle \tau_y \rangle$ of either sign, which vanishes at long times.
Previous work has shown, using effective rate equations for single-site models, that 
chiral driving of phonons can induce effective fields for dipolar $4f$ magnets  \cite{juraschek2022giant,Intro_ChiralPhonon_chaudhary2023giant,luo2023large}. 
Our work provides a nontrivial generalization of this idea to multipolar orders on the lattice. 
To understand this physics, 
we note that the phonons couple linearly to the quadrupoles, motivating 
a proxy Hamiltonian 
$H^{\rm sp}_{\rm drive} = -A_{\rm eff} \sum_\br (\tau_{\br x} \cos(\Omega t) + \tau_{\br z} \cos(\Omega t - \phi))$
which
directly drives quadrupoles.
The effective field $A_{\rm eff} = \lambda Q_0$ with $Q_0$ being the peak phonon amplitude (which
depends on the drive strength $A$).
We study this using a Floquet-Magnus expansion \cite{suppmat}
since $\Omega=\hbar\omega/J_0 \gg 1$, i.e. the phonon dynamics is much
faster than pseudospins \cite{Intro_Floquet_Oka_ARCMP2019,bukov2015universal,eckardt2015high}. In standard notation,
\begin{eqnarray}
\! H^{\rm sp}_{\rm drive,0}&=&0; ~~~H^{\rm sp}_{\rm drive,\pm}\! = -\frac{\lambda Q_0}{2} \! \sum_\br (\tau_{\br x} \!+\! e^{\mp i\phi} \tau_{\br z}) \\
\!\!\! H^{\rm sp,eff}_{\rm drive} \! &=& \frac{1}{\Omega} [H^{\rm sp}_{\rm drive,+}, H^{\rm sp}_{\rm drive,-}] \!= \frac{\lambda^2 Q_0^2}{\Omega} \sin\phi \!
\sum_\br \tau_{\br y}
\end{eqnarray}
This phonon-induced effective field in $H^{\rm sp,eff}_{\rm drive}$, $(-\lambda^2 Q_0^2/\Omega) \sin\phi$, explains 
the induced $\phi$-dependent octupolar order for $T \!>\! T_{\rm oct}$ shown in Fig.~\ref{fig:nk2phonon}(a).

\noindent {\it Dynamical switching of octupolar order.---}
We next ask if this pulsed phonon drive can also be used to switch the sign of octupolar order for $T \!<\! T_{\rm oct}$. Fig.~\ref{fig:nk2phonon}(b)
shows the impact of the two-phonon coherent drive for $\phi=\pi/2$ at $T/T_{\rm oct}=0.8$. Using a drive which
leads to a peak phonon amplitude $Q_0 \!\sim \! 1$-$3$ (i.e., $\sim\! 1$-$3\%$ atomic displacement), we find that octupolar order in the positive domain is suppressed
or can even switch sign, depending on the drive amplitude
while the octupolar order in negative domains (not shown) is nearly unaffected.
Furthermore, the slow evolution of 
$\la \tau_y\ra$ at late
times indicate that heating effect of pseudospins is weak at these
timescales, although additional cooling might be needed in experiments
to extract the energy dissipated by the phonon modes.
Nonlinear phononics experiments on CoF$_2$ have realized
large atomic displacements $\sim 2\%$ of Raman-active modes by driving an infrared-active phonon with $500$ fs
pulse of 10MV/cm electric field \cite{Intro_OpticalCEF_Cavalleri_NatPhy2020}; such methods could potentially 
be extended to the osmates explored in this work.

\begin{figure}[t]
\centering
\includegraphics[width=0.45\textwidth,height=0.45\textwidth]{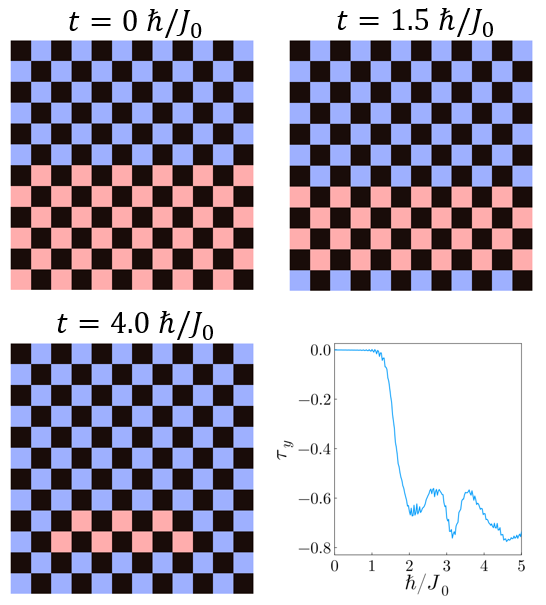}
\caption{Visualization of octupolar order switching for an initial two-domain configuration (assuming periodic boundary conditions). 
We drive the $Q_x$ and $Q_z$ phonons with a relative phase $\phi\!=\!\pi/2$ for a short time $t=3\hbar/J_0$ ($\sim 1$\,ps) and show 
configurations at representative times. Red and blue squares respectively depict pseudospins with $\langle \tau_y \rangle>0$ and
$\langle \tau_y \rangle<0$. We clearly see the shrinking of the positive (red) domain and growth of the negative (blue) domain.
This domain evolution is also reflected in the net octupolar magnetization $\langle \tau_y \rangle$ averaged over the whole lattice 
(bottom right panel) which is initially zero but evolves to negative values.}
\label{fig:configsPics}
\end{figure}

We find that octupolar switching is mediated by flipping of pseudospins at domain wall boundaries by the
phonon-induced field $(-\lambda^2 Q_0^2/\Omega) \sin\phi$; for $\phi\!=\!\pi/2$, this promotes growth of $\la \tau_y \ra\!<\!0$
domains.
In Fig.\ref{fig:configsPics}, we depict this time evolution of pseudospins under a two-phonon drive 
starting with an initial 3D domain wall configuration with periodic boundary conditions, and a thermal population of phonon momenta. We show a
single 2D slice through the lattice, with colours indicating positive $\langle \tau_y \rangle$ (red) and negative 
$\langle \tau_y \rangle$ (blue) and the black sites denoting the non-magnetic sublattice. 
One can clearly see the two-phonon drive inducing a shrinking of the $+ \langle \tau_y \rangle$ domain as a function of time.
The bottom right panel of Fig.\ref{fig:configsPics} shows $\la \tau_y\ra$ averaged over the full lattice, 
evincing growth of octupolar magnetization; we attribute the superposed rapid oscillations to the high-frequency phonons.
Additional data and an effective
model of this mechanism is given in the SM \cite{suppmat}.


Recent remarkable experiments \cite{torre2021mirror} have shown that one can unambiguously diagnose the presence and, significantly, \emph{sign} of 
ferro-octupolar order using
rotational anisotropy second harmonic generation (SHG). The octupolar order is characterized through interference between the time reversal even and odd channels, with the sign of the octupolar order determining the positions of the measured intensity maxima \cite{torre2021mirror}. 
We propose a tr-SHG  experiment which compares the position of intensity maxima of equilibrium versus the driven system
 at time $t\sim 3.0 \hbar/J_0$ ($\sim 2.0 \rm ps$)  to diagnose the theoretically predicted
build-up and switching of octupolar order. Additional plots showing dependence of our results on the phonon damping
parameter $\eta$ are given in the SM \cite{suppmat}.

\noindent {\it Summary.---}
We have developed computational tools to simulate the equilibrium phases and nonequilibrium
dynamics of multipolar degrees of freedom
coupled to dissipative Einstein phonons in multipolar Mott insulators.
Our equilibrium phase diagram for the model hosts ferro-octupolar and quadrupolar orders at low
temperature, as partly described by previous work, but
additionally shows that the quadrupolar order is accompanied by weak lattice displacements.
We have also shown how pump-probe techniques and two-mode phonon drive can
uncover unique dynamical signatures and enable the ultrafast control of the
ferro-octupolar state. This study opens up distinct directions in the detection and control of complex hidden orders
in quantum materials. 

This research was funded by an NSERC Discovery Grant (AP),
an Ontario Graduate Scholarship (KH), and an NSERC CGS-D fellowship (RS).
We thank Sreekar Voleti, David Hsieh, Prashant Padmanabhan,
Emily Zhang, William Bateman-Hemphill, and Cristian Batista 
for extremely helpful discussions.
Numerical computations were performed on the Niagara supercomputer at the SciNet HPC 
Consortium and the Digital Research Alliance of Canada. Data for the figures in this work is available, see \cite{dataAvail}.

\bibliography{main}

\section*{End Matter}

\subsection*{Double Perovskite Crystal Structure}
\begin{figure}[h]
\centering
\includegraphics[width=0.45\textwidth]{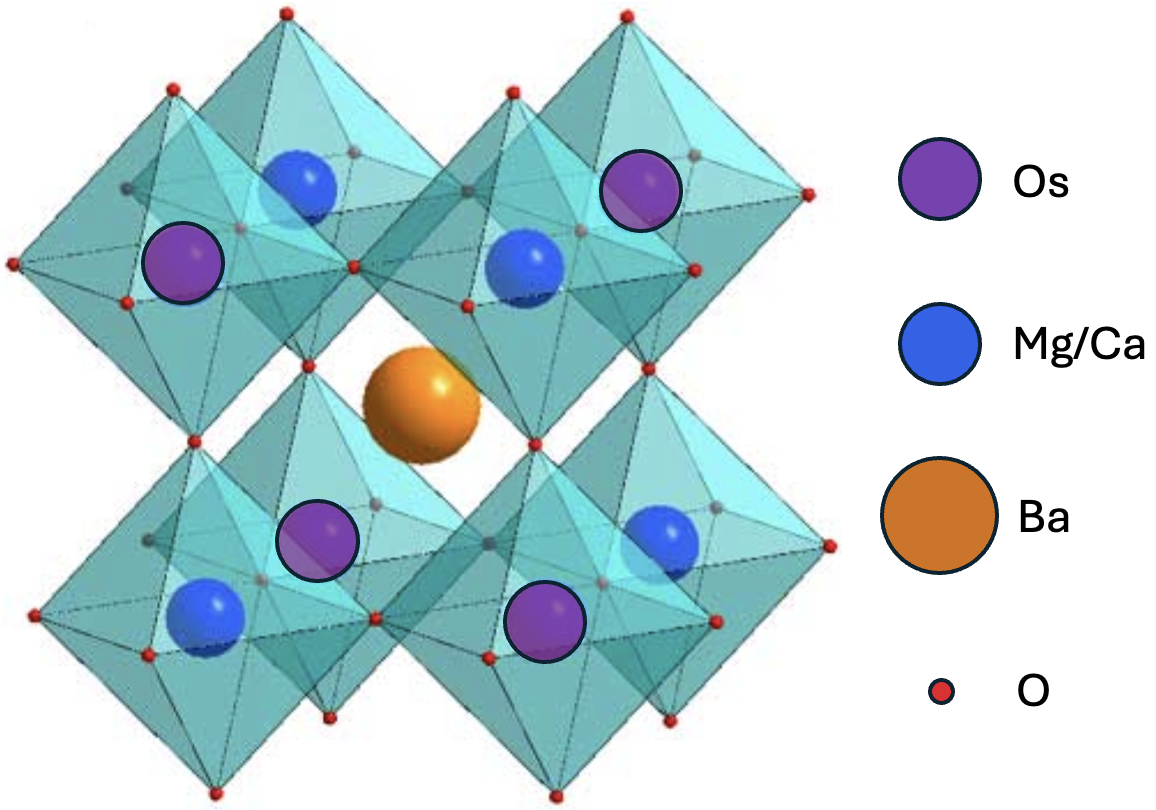}
\caption{Crystal structure for the double perovskite system discussed in this work. Magnetic osmium ions (purple) sit at the
centers of octahedral oxygen cages, and the lattice of Os ions forms an fcc lattice.}
\label{fig:crystalStructure}
\end{figure}
The ordered double perovskites Ba$_2${$M$}OsO$_6$ with $M$=Mg,Ca have a cubic $Fm{\bar3}m$ structure as depicted in Fig.~\ref{fig:crystalStructure},
with magnetic Os and non-magnetic Mg/Ca each encased in an octahedral oxygen cage and forming two interpenetrating fcc lattices. Exciting the
$E_g$ phonon modes of the octahedra is shown to be a route to study octupolar order on the Os sites.

\bigskip

\subsection*{Details of the Pump-Probe Simulations}
\label{app:Simulations}
In this section, we provide a discussion of the Monte Carlo and molecular dynamics simulations conducted in the `Detection of octupolar order' section. To simulate pump-probe dynamics, we
use equilibriated Monte Carlo configurations of the spins and phonon coordinates as initial conditions, and excite the
$Q_x$ phonon by choosing its momenta from a {\it displaced}
Maxwell thermal distribution. In doing so, we model the pump-beam as a spatially uniform impulsive momentum kick to the phonon 
at all sites \cite{ning2023coherent}; we check that our quench leads to a maximum normalized phonon displacement $Q_\alpha \lesssim 1$
over the time evolution. With these initial conditions,
we compute the Schr\"odinger evolution for the pseudospin wavefunction \cite{dahlbom2022geometric}
together with Newton's laws for the classical phonons,
and track time evolution of observables averaged over a large number of initial Monte Carlo configurations out to times
$t=15 \hbar/J_0$ ($\approx 10$\,ps) for system sizes up to $10^3$ sites. We have included a weak
damping of the ${\cal E}_g$ phonons, incorporating an additional term $(-\eta P_\alpha)$ in the force equation $d P_\alpha/dt$,
to capture energy dissipation to the bath of acoustic phonon modes. 
This mechanism of energy damping via phonons is more relevant to Mott insulators than Gilbert damping of spin modes
which plays a role in metallic magnets.

\end{document}


\preprint{APS/123-QED}
\title{Supplementary Material: \\ Phonon-driven multipolar dynamics in a spin-orbit coupled Mott insulator}
\author{Kathleen Hart}
\thanks{These authors contributed equally to this work.}
\affiliation{Department of Physics, University of Toronto, 60 St. George Street, Toronto, ON, M5S 1A7 Canada}
\author{Ruairidh Sutcliffe}
\thanks{These authors contributed equally to this work.}
\affiliation{Department of Physics, University of Toronto, 60 St. George Street, Toronto, ON, M5S 1A7 Canada}
\author{Gil Refael}
\affiliation{Department of Physics, California Institute of Technology, Pasadena CA 91125, USA}
\affiliation{Institute for Quantum Information and Matter, California Institute of Technology, Pasadena CA 91125, USA}
\author{Arun Paramekanti}
\email{arun.paramekanti@utoronto.ca}
\affiliation{Department of Physics, University of Toronto, 60 St. George Street, Toronto, ON, M5S 1A7 Canada}

\date{\today}

\maketitle

\beginsupplement

\section{Short time dynamics of phonons in the pump-probe setup}
\label{app:short-time}
In the main text, Fig.~1(b), we have shown the energy exchange between the $Q_x$ and $Q_z$ phonons enabled by the presence of octupolar order up to time $t \sim 15 \hbar/J_0$ which masks the rapid oscillations of the phonons at time scales corresponding to the phonon period. To highlight these rapid oscillations, we show below the short-time dynamics of the two phonon modes.
\begin{figure*}[h]
\centering
\includegraphics[width=0.8\textwidth]{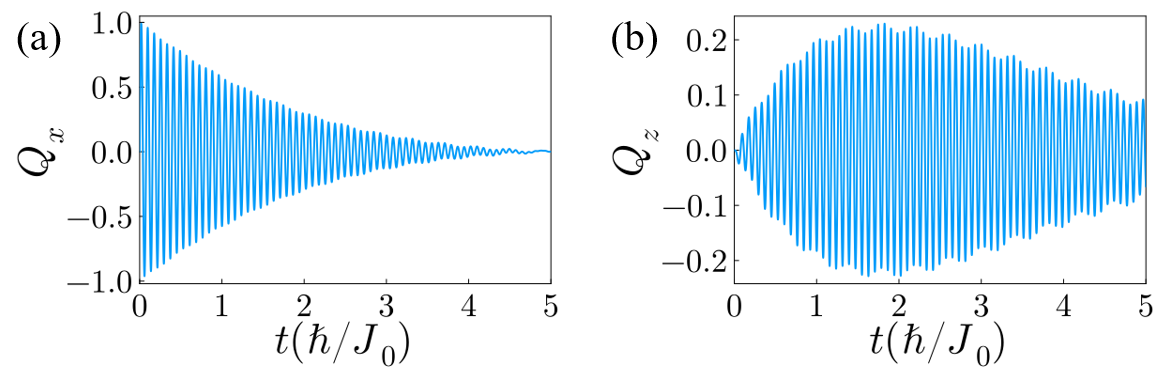}
\caption{Short time phonon dynamics reveals the rapid oscillations of Fig 1.(b) in the main text, averaged over 80 configurations for (a) $Q_x$ phonon and (b) $Q_z$ phonon.}
\label{fig:phonons_octupolar_full}
\end{figure*}

\section{Mean field model of coupled pseudospin-phonon dynamics}
\label{app:mean-field}

\subsubsection{Pseudospin dynamics}

In this section, we consider a single spin in a ferro-octupolar ordered state with a phonon mode $Q_x$ coupled to $\tau_x$. We construct such a model in order to simulate the effect of a single phonon pump in the pseudospin-1/2 model as a potential mechanism by which the phonon cross-amplitude, shown in Fig. 1(b) of the main text, is induced. Here, we model the effect of nearest-neighbour pseudospin interactions as a constant octupolar mean field $h$ 
pointing along $\tau_y$, and the pumped phonon as a sinusoidal time-dependent drive. Thus, the single spin Hamiltonian is:
\begin{equation}
    H = -h\tau_y - A_{\rm eff} \sin(\Omega t + \varphi) \tau_x,
\end{equation}
where $A_{\rm eff}=\lambda Q_0$ encompasses the spin-phonon coupling strength $\lambda$ and the phonon amplitude $Q_0$.
The Heisenberg equations of motion may thus be written as:
\begin{eqnarray}
    \dot{\tau}_x &=& -h\tau_z\\
    \dot{\tau}_y &=& A_{\rm eff} \sin(\Omega t + \varphi)\tau_z\\
    \dot{\tau}_z &=& h\tau_x - A_{\rm eff} \sin(\Omega t + \varphi) \tau_y.
\end{eqnarray}
Given that fluctuations away from the ferro-octupolar ordering are small, we make the simplifying approximation that $\tau_y\simeq1$ and is constant in time. Therefore, the equations of motion reduce to:
\begin{eqnarray}
    \dot{\tau}_x &=& -h\tau_z\\
    \dot{\tau}_z &=& h\tau_x - A_{\rm eff} \sin(\Omega t + \varphi).
\end{eqnarray}
Solving these coupled equations with the initial conditions $\tau_x(0)=\tau_x^0$ and $\tau_z(0)=\tau_z^0$, we find
\begin{equation}
\label{eq:singleDrivePseudoModelx}
    \tau_x(t) = \left(\tau_x^0+\frac{A_{\rm eff} h \sin\varphi}{\Omega^2-h^2}\right) \cos (ht)
    + \left(-\tau_z^0+\frac{A_{\rm eff} \Omega \cos(\varphi)}{\Omega^2-h^2}\right)
    \sin(ht)
    - \frac{A_{\rm eff} h}{\Omega^2-h^2}\sin(\Omega t+\varphi)
\end{equation}
\begin{equation}
    \label{eq:singleDrivePseudoModelz}
    \tau_z(t) = \left(\tau^0_x+\frac{A_{\rm eff} h \sin(\varphi)}{\Omega^2-h^2}\right) \sin(ht)
    +\left(\tau_z^0-\frac{A_{\rm eff} \Omega \cos(\varphi)}{\Omega^2-h^2}\right) \cos(ht)
    + \frac{A_{\rm eff} \Omega}{\Omega^2-h^2}\cos(\Omega t+\varphi)
\end{equation}

\begin{figure*}[h]
\centering
\includegraphics[width=0.8\textwidth]{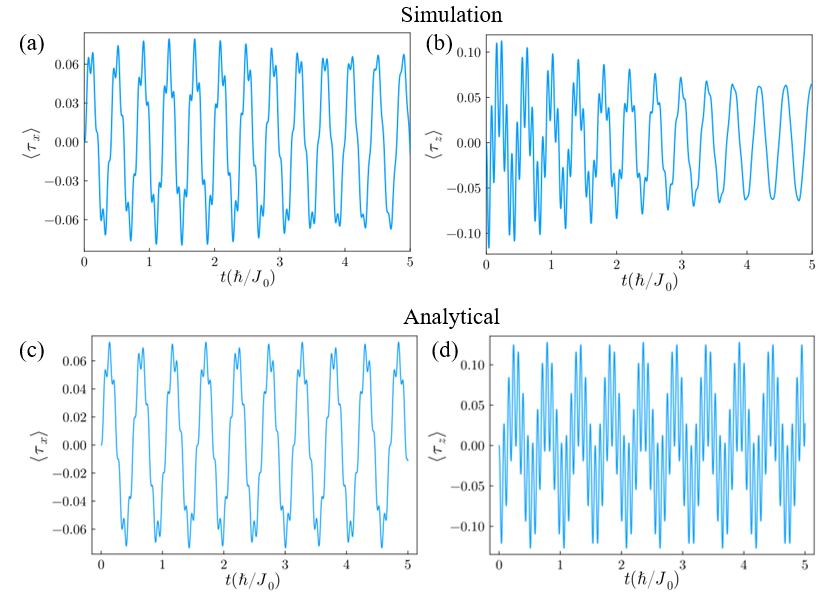}
\caption{(a),(b) Spin dynamics in the ferro-octupolar phase from the full lattice Monte Carlo and molecular dynamics simulations
corresponding to Fig.~1(b) of main manuscript ($T < T_{\rm oct}$). We observe small, coherent rapid oscillation in the (a) $\tau_x$ and (b) $\tau_z$ components which track 
the phonons to which they are coupled, in addition to slower oscillation which correspond to the timescale governed by 
$\hbar/J_0$.
(c),(d) Analytical solution to the coupled differential equations given by Eqs. \ref{eq:singleDrivePseudoModelx}, shown in (c), and \ref{eq:singleDrivePseudoModelz}, shown in (d). We use parameters as in the numerical pump-probe simulations, namely $A_{\rm eff}=5$, $\Omega=80$, $h=12$, and $\phi=0$, with initial conditions $\tau^0_x = \tau^0_z = 0$.}
\label{fig:toydynamics}
\end{figure*}

Figs.~\ref{fig:toydynamics}(c) and (d) plots the dynamics in Eqs.~\ref{eq:singleDrivePseudoModelx} and \ref{eq:singleDrivePseudoModelz}, and it shows
two key features: a slower oscillation which comes from the precession of the spins about the local magnetic field (on timescale $\sim 0.5\hbar/J_0$), on which is superposed a rapid oscillation that results from the linear coupling of the fast phonon coordinate to the spin components 
(on timescale $\sim 0.1\hbar / J_0$). 

To compare the mean field model to our numerical simulations, we present plots showing the accompanying spin dynamics for the pump-probe simulations shown in Fig. 1.(b) of the main text.
Figs. \ref{fig:toydynamics}(a) and (b) in the ferro-octupolar phase (for $T < T_{\rm oct}$) show the corresponding
dynamics of the pseudospin components $\tau_x, \tau_z$ averaged over the lattice and over $\sim 80$ 
initial configurations where $\la \tau_y \ra > 0$. While $\tau_y \approx 1$ (not shown), there
are rapid oscillations in $\tau_x,\tau_z$ which follow the phonon frequency,
as well as slower oscillations which arise from spin precession around $\tau_y$ direction as clearly seen from the envelope. This toy model is clearly able to capture generally the amplitude and frequency of the oscillations seen in both $\tau_x$ and $\tau_z$ in the pump-probe simulations.

 \begin{figure*}[h]
\centering
\includegraphics[width=0.8\textwidth]{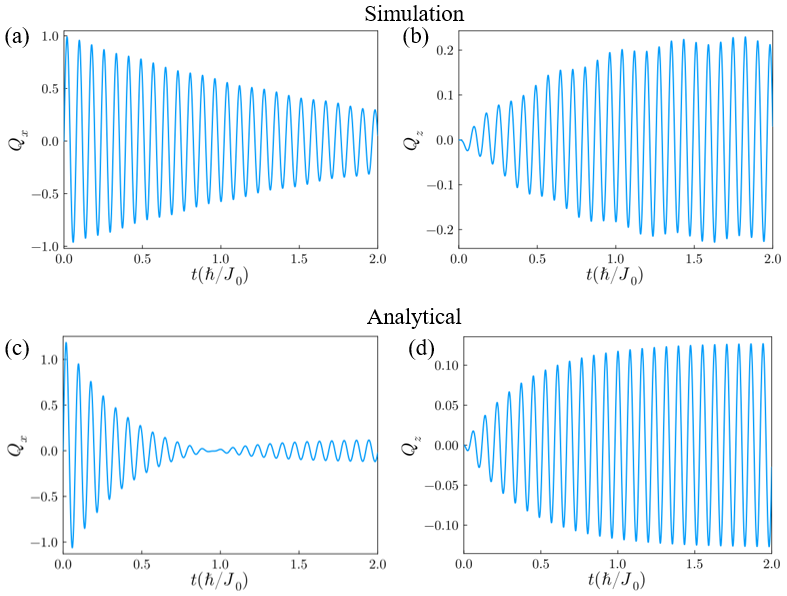}
\caption{(a), (b) Short time phonon dynamics from the full MC and molecular dynamics simulations as shown in the main text. (c),(d) Results for the phonon dynamics from the analytical solution of the mean field pseudospin-phonon model showing that it is in reasonable agreement with the full simulation results shown in panels (a),(b).}
\label{fig:phonons_octupolar}
\end{figure*}

\subsubsection{Phonon dynamics}

Using the previous section as a starting point, we may compute the back-action on the phonon degrees of freedom to capture 
the phonon dynamics in the ferro-octupolar phase shown in Fig. \ref{fig:phonons_octupolar}(a) and (b). Explicitly, if we treat the spin equations of motion shown in Eqs. \ref{eq:singleDrivePseudoModelx} and \ref{eq:singleDrivePseudoModelz} as a driving term through the spin-phonon coupling, and note that the dominant effect of this drive will be due to the terms resonant with the phonon frequency $\Omega$, we may reduce this problem to that of a driven damped harmonic oscillator. Explicitly, one can write:
\begin{equation}
\frac{dQ_x}{dt}=P_x/M
\end{equation}
\begin{equation}
\frac{dQ_z}{dt}=P_z/M
\end{equation}
\begin{equation}
\label{eq:SM-phononEqsMotion-X}
\frac{dP_x}{dt}=-M\Omega^2Q_x+\frac{\lambda A_{\rm{{eff}}}h}{\Omega^2-h^2}\sin(\Omega t) -\eta P_x
\end{equation}
\begin{equation}
\label{eq:SM-phononEqsMotion-z}
\frac{dP_z}{dt}=-M\Omega^2Q_z-\frac{\lambda A_{\rm{{eff}}}\Omega}{\Omega^2-h^2}\cos(\Omega t) -\eta P_z
\end{equation}
where the coefficients are chosen as discussed in the main text.
Thus, we may then write the trajectories of the $Q_x$ and $Q_z$ phonons as:
\begin{equation}
\label{eq:singleDrivePseudoModel}
Q_\alpha(t)=D_\alpha e^{-\eta t/2}\sin(\Omega't+\Phi_h^\alpha) +C_\alpha\cos(\Omega t-\Phi)
\end{equation}
using the initial conditions $Q_{\alpha}(0)=0.0,0.0$ and $P_\alpha(0)=1.0,0.0$ for the $Q_x$ and $Q_z$ phonons, respectively. Here, $\Phi=\pi/2-\phi_D$, $\Omega'=\sqrt{\Omega^2-(\eta/2)^2}$, $\Phi^\alpha_h=\arctan(\frac{\Omega'C_\alpha\cos(\Phi)}{P_0/M+(\eta/2)(C_\alpha\cos(\Phi))-C_\alpha\Omega\sin(\Phi)})$, $C_\alpha=F^\alpha_0/\eta M\Omega$, and $D_\alpha=C_\alpha\cos(\Phi)/\sin(\Phi_h^\alpha)$. $F^\alpha_0$ are the driving amplitudes shown in Eqs. \ref{eq:SM-phononEqsMotion-X} and \ref{eq:SM-phononEqsMotion-z}, and $\phi_D=\pi/2,0$ for $Q_x$ and $Q_z$ respectively. The analytical results for 
 $Q_x$ and $Q_z$ are shown in Fig. \ref{fig:phonons_octupolar}(c) and (d), respectively, and qualitatively agree with the dynamics seen in the full lattice model.

 \section{The lab frame Floquet-Magnus expansion}
\label{app:magnusExpansion}
In this section, we provide a description of the lab frame Floquet-Magnus expansion used in this work. There has been significant recent theoretical and experimental interest in \emph{time-periodic} systems and Hamiltonians of which this work is but one example \cite{liu2018floquet,junk2020floquet,nuske2020floquet}. To describe such systems, it has become commonplace to use Floquet theory, decomposing the dynamics of the system into a product of unitary operators describing dynamics within the period of the system and longer time dynamics, extending over multiple periods \cite{eckardt2015high}. Although Floquet theory provides a neat description of time-periodic Hamiltonian systems, in practice, computation of the operators governing the dynamics of the system (the stroboscopic kick operator for short-time dynamics and the effective Hamiltonian for long-time dynamics) can be challenging, or impossible, to compute exactly. Thus, in systems where the drive frequency, $\Omega$, is significantly higher than other energy scales in the problem, one can employ an expansion in powers of $1/\Omega$ to calculate the effective Hamiltonian. 

To employ the Floquet-Magnus expansion, we begin with a decomposition of the Hamiltonian into its spectral components, $H_k$, where one may define:
$$
H(t)=\sum_{k=-\infty}^\infty H_k e^{ik\omega t}.
$$
Having calculated the relevant $H_k$, a computation of $H_{\rm{eff}}$, where $H_{\rm{eff}}=\sum_{j}H^{(j)}_{\rm{eff}}$ is possible. Explicitly, for this work, we computed $H_{\rm{eff}}$ to leading order only, where: 
\begin{equation}
H_{\rm{eff}}\simeq H^{(1)}_{\rm{eff}}=\frac{1}{\Omega} \left[H_+,H_{-}\right]
\end{equation}
such that $k=+1$ corresponds to $H_+$ and $k=-1$ corresponds to $H_-$. In the main text, we compute the Magnus expansion on $H_{\rm drive}^{\rm sp}$.

Although, in principle, it can be useful to choose a stroboscopic kick operator in advance of computing the effective Hamiltonian, it is tantamount to a change of reference frame not necessary in the large frequency limit, and we have thus elected not to do so in this work. In practice we use this `lab-frame' expansion to describe the physics near zero energy, including the generation of an effective magnetic field through a two-phonon drive.

\section{Mean field model of octupolar switching dynamics near a domain wall}

\begin{figure}[h]
\centering
\includegraphics[width=0.4\textwidth]{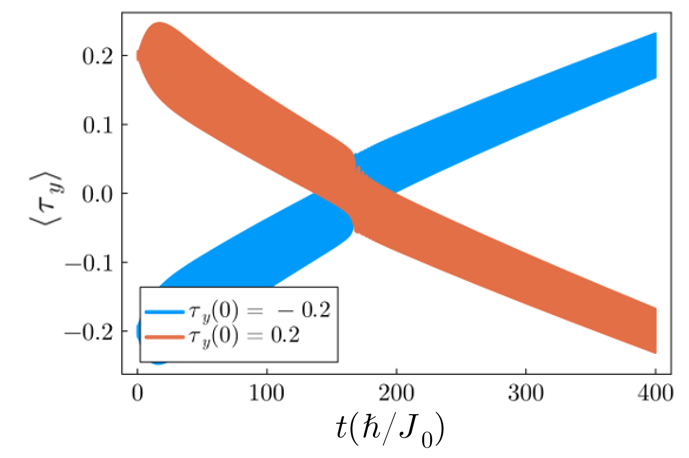}
\caption{Toy model of a driven spin, starting near the $xz$ plane, with a time dependent field in $\tau_y$ and a two phonon sinusoidal drive. Here, the phonons are driven $\pi/2$ out of phase and coupled to the $\tau_x$ and $\tau_z$ spin components according to Eq.\ref{eq:modelSpinFlipHam}. Spin trajectories for the state $\langle\tau_y(0)\rangle=+0.2$ was driven with $\phi=+\pi/2$, while the state with $\langle\tau_y(0)\rangle=-0.2$ was driven with $\phi=-\pi/2$. This toy model confirms that spins near, or at, domain walls are able to be flipped as a result of the phonon-induced effective field.}
\label{fig:spinFlipToyModel}
\end{figure}

In this section, we seek to capture the mechanism for octupolar order switching at domain walls using a simplified model of the pseudospin-1/2 system. We begin by simulating spins sitting at or near domain walls using a mean field description. Explicitly, we do this by initializing the spin of interest close to the $xz$ plane in addition to assuming that the effective field produced by neighbouring spins sitting on either side of the domain wall will cancel out. Thus, the only contributors to the effective field in $\tau_y$ are neighboring spins which similarly sit on the domain wall and each contribute an effective field approximately equal to the $\tau_y$ component of the spin of interest. We may thus write our Hamiltonian as:
\begin{equation}
\label{eq:modelSpinFlipHam}
H=-h(t)\tau_y-A_{\rm eff}\cos(\Omega t)\tau_x-A_{\rm eff}\cos(\Omega t-\phi)\tau_z,
\end{equation}
where $h(t)\propto \langle \tau_y\rangle(t)$, where $\langle \tau_y\rangle(t)$ represents the expectation value of $\tau_y$ for the single spin which we are simulating, $\Omega$ is the phonon frequency,
and $A_{\rm eff}=\lambda Q_0$ with $Q_0$ being the phonon amplitude and $\lambda$ being the spin-phonon coupling. 
We then simulated two near-domain wall conditions, explicitly, looking at both sides of a domain wall with $\langle \tau_y \rangle >0$ and $\langle \tau_y \rangle <0$. These spins were then driven using a two phonon drive as described in Eq.\ref{eq:modelSpinFlipHam}, with phonon parameters matching those used in the full model ($A_{\rm eff}\sim 5$, $\Omega\sim 80$) and $\phi=\pm \pi/2$ depending on the initial state to induce octupolar order switching in both cases. The results are shown in Figure \ref{fig:spinFlipToyModel}, where we observe a similar switching effect to that seen in the full numerical model of the lattice, albeit at a different timescale.



\section{Dependence of dynamical effects on the phonon damping constant}

\subsubsection{Pump-probe dynamics}

\begin{figure*}[h]
\centering
\includegraphics[width=0.7\textwidth]{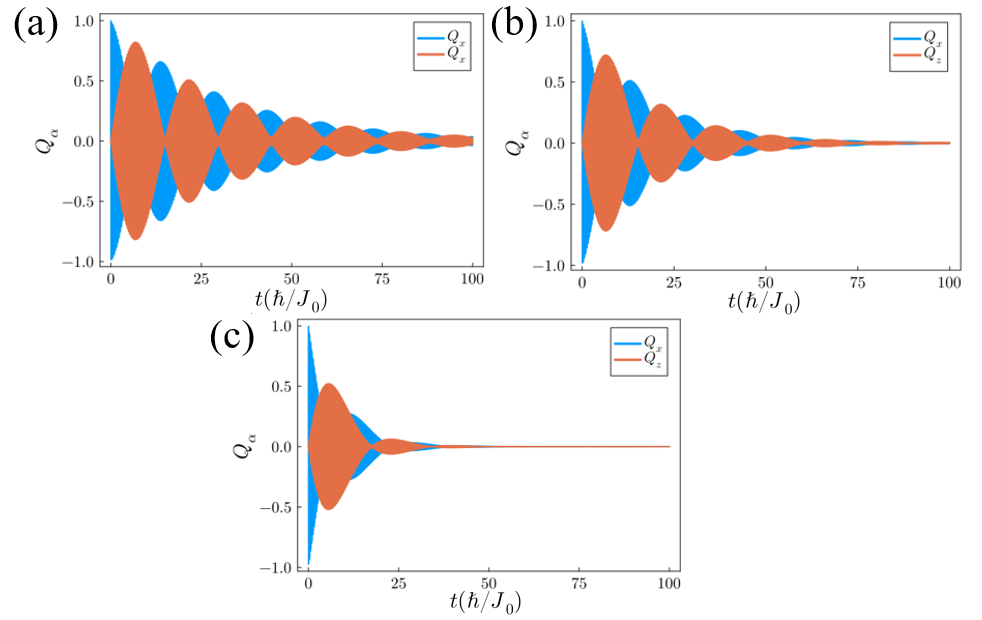}
\caption{Phonon dynamics showing the phonon-coordinates, $Q_x, Q_z$, for simulations of pump-probe experiments. Here, we model the pump pulse as a impulsive kick to the $Q_x$ phonon at time $t=0$, and observe coherent oscillation between the two phonons at timescales characteristic of the spin exchange energy, with rapid oscillations too small to be seen on these timescales. Here, we present data for $T/J_0=2.1$ at
three damping strengths $\eta$ which are smaller than the values used in the main
text. Explicitly, we show (a) $2\pi\eta/\omega=0.007$, (b) $2\pi\eta/\omega=0.013$, and (c) $2\pi\eta/\omega=0.035$.}
\label{fig:pumpProbe-newDamping}
\end{figure*}

As introduced in the main text, we include a phenomenological damping in the equations of motion for the phonons, to account for the dissipation of energy seen in real materials. Such damping depends on details of the physical system, and we highlight the impact thereof on pump-probe experiments by implementing simulations at $T/J_0=2.1$ while varying  damping constant $\eta$, {choosing values smaller than that used in the main text}.  In Fig.\ref{fig:pumpProbe-newDamping}, we present data for pump-probe simulations with damping constants {$2\pi\eta/\omega=0.007, 0.013, 0.035$}. In addition to observing the expected increase in rate at which energy is damped out of the system, we observe, particularly at the large damping $2\pi\eta/\omega=0.035$, a change in the shape of the envelope of the long timescale oscillations which transfer energy between the $Q_x$ and $Q_z$ phonons.

\section{Octupolar order switching under continuous-wave driving }

\begin{figure}[h]
\centering
\includegraphics[width=0.95\textwidth]{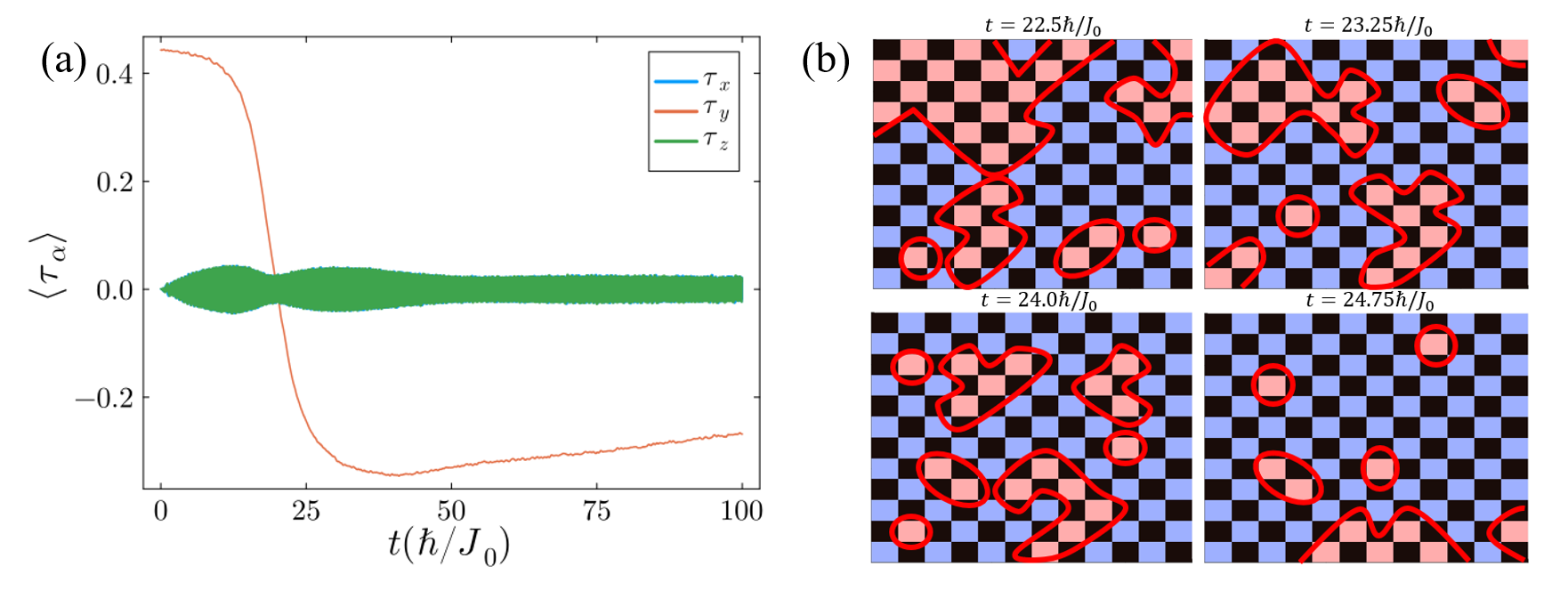}
\caption{Spin dynamics in the presence of a continuous-wave phonon drive in the octupolar-ordered phase for $2\pi\eta/\omega \approx 0.007$, averaged over only $+\langle \tau_y \rangle$ domains out of 160 MC configurations. (a) For small damping the $\tau_y$ component of the spin switches sign as the phonon coordinate reaches a maximum value $Q_\alpha \sim 1.5a$. (b) Two dimensional slice through a single configuration for the averaged dynamics shown in (a). One can see, as in Fig.3 of the main text, that the switching predominantly occurs at the domain walls (highlighted in red).}
\label{fig:drivenPhonons_pseudo}
\end{figure}

{Having established a framework for octupolar order switching in the pseudospin model, we move to a discussion of the effect of the drive form on said switching. Specifically, in this section, although less experimentally feasible than Fig.2.(b) in the main text, we show the effect of a continuous-wave, two phonon-drive on a ferro-octupolar ordered state.} Here, we select only positive  $\langle \tau_y \rangle$ configurations and fix the drive amplitude, $A=0.2$, and temperature $T/J_0=3.0$, while varying $\eta$. Explicitly, in Fig. \ref{fig:drivenPhonons_pseudo}, we show the spin trajectories, averaged over  $\sim 100$ configurations with $+\langle \tau_y \rangle$ for {$2\pi\eta/\omega \approx 0.007$.} For $2\pi\eta/\omega=0.007$ we observe the average phonon coordinates saturate to a steady state with amplitude $Q_\alpha \sim 1.5$, which corresponds to a $1.5\%$ atomic displacement.


{In Fig.\ref{fig:drivenPhonons_pseudo}.(b) we provide further pictorial evidence for octupolar order switching occurring predominantly at domain walls. In Fig. 3 of the main text we show a two-dimensional slice of an illustrative initial spin configuration having \emph{only} one domain wall in the presence of a two-phonon drive. Here, we present an equivalent figure, highlighting the behaviour of a two-dimensional slice through a configuration used to produce Fig.\ref{fig:drivenPhonons_pseudo}.(a). We use red lines to highlight the $+ \langle \tau_y \rangle$ domains. The shrinking of these domains, similarly to that shown in Fig. 3 of the main text, gives evidence that the octupolar order switching is enabled by the effective magnetic fields produced by the phonons flipping spins at the domain boundary.}

\bibliography{main}